\documentstyle[12pt,aasms4]{article}
\begin{document}

\title{Host galaxies of 2MASS-selected QSOs at redshift over 0.3
\footnote{Based on observations obtained with the Canada France Hawaii
Telescope, which is operated by NRC of Canada, CNRS of France, and the
University of Hawaii.}}

\author{J.B.Hutchings and A. Cherniawsky}

\affil{Herzberg Institute of Astrophysics, 5071 West Saanich Rd.
Victoria, B.C. V9E 2E7, Canada; john.hutchings@nrc.ca}

\author{R.M.Cutri and B.O.Nelson}
\affil{IPAC, MS 100-22,CalTech, 770 S Wilson Ave, Pasadena, CA 91125}

\begin{abstract}

We have obtained optical imaging with the Canada France Hawaii Telescope
(CFHT) of 21 2MASS-selected QSOs of redshift greater than 0.3. This paper
complements the sample of lower redshift 2MASS QSOs previously published. 
The QSOs have higher overall and nuclear luminosity, bluer colours, and 
higher ratio of nuclear to host flux than the lower redshift sample. 
From these and other properties, we argue that the sample is consistent 
with the emergence of the AGN from dusty starbursts following major tidal
interactions between galaxies.

\end{abstract}

\keywords{galaxies:quasars}

\section{Introduction and observations}

This paper is a sequel to the paper by Hutchings, Maddox, Cutri, and
Nelson (2003: paper 1), which presented optical imaging of QSOs of redshift
0.3 and lower, selected from the Two Micron All Sky Survey (2MASS). HST
imaging of a different sample of 2MASS QSOs is given by Marble et al (2003).
In this paper we present optical imaging of a sample of 2MASS QSOs at 
redshifts above 0.3. The sample of QSOs is the same as in paper 1, but 
with the higher redshift bin.

The QSOs are identified by spectroscopy after selection by colours
J-K larger than 2.0, and are classified as types 1, 2 or intermediate,
based on the ratio of narrow to broad line emission. These do not include
any previously known AGN. 2MASS QSO selection is described in Cutri et al.
(2002).  The present study includes objects between
-30$^o$ and 60$^o$ declination, and thus suitable for observation with the 
CFHT. Paper I included data on 76 of 243 eligible objects, and this paper
includes data on 21 of 64 eligible
objects. All but 2 of the eligible objects have redshifts below 0.7.

The observations were taken as service-observing `snapshots' with
the Megaprime camera and r-band filter, with exposures of 340 secs. The pixel
sampling is 0.187 arcsec. (The paper 1 data were taken with R band, and 0.206
arcsec sampling, and 200 sec exposures, with the CFHT 12K camera. The CFHT
website gives details of the different filters.) All
objects had two independent images, 5 had four images, and two had 8 images.
The data were obtained over the period April through June 2004, in seeing
conditions that varied from 0.6 to 1.3 arcsec FWHM. Table 1 lists the
objects observed, and various of their properties, from the 2MASS database
and also measures from the present work.

As the selection of observed objects was dictated mainly by scheduling,
we show in Figure 1 the distribution of observed objects within the whole
sample of 64 high redshift objects, and also include the lower redshift
sample from paper 1.
The observed subsample is slightly skewed to higher luminosity, but there
is K-band luminosity overlap of about 15 new objects with 21 of the highest
luminosity objects from the lower redshift sample. In terms of J-K colour,
the observed subset is slightly
skewed to redder objects than the average by about 0.5 magnitudes.
The spectral classifications among the new sample
are representative of the 2MASS QSOs in this redshift range.

\section{Data processing and measurements}

After removal of the instrumental signature, subimages were generated of
each QSO and 3 PSF stars from the same CCD and observation, as near as
possible to the
QSO. The two (or more) images of each field were combined, to help remove
cosmic rays and to increase the signal. In a couple of cases, some of
the images had very different FWHM as they were taken on different nights,
and such images were not combined.

PSF stars were chosen to have more signal than the QSOs, and to be
free of nearby companions and image flaws. Any other stars near the PSF
stars were edited out of the images before generating the profiles, to
get the most representative PSF profiles. The images and luminosity
profiles were compared among the PSF stars, and generally the cleanest
one was picked as the PSF to be used. Occasionally
two PSF stars were co-added, to increase the signal, after checking that
the profiles agreed.

Measurements were made of every individual image, as well as the
combined images, to give independent values of the fluxes and PSF-removal.
The measurements made were the flux of the QSO and the PSF stars, and the
image FWHM from profiles. Profiles were generated by the IRAF task ellipse,
after careful sky removal. Companions near the QSOs were included in
their profiles,
although in some cases, well separated companions were edited out to better
study the extended flux centered on the QSO. Figure 2 shows profiles
from one object as an example.

The host galaxy fluxes were estimated in two ways. First, the flux difference
was measured between the QSO and PSF profiles, after scaling the PSF
profile to the same peak value as the QSO. This yields a measure
of the resolved flux which is a lower limit to the QSO host flux. Second,
differently scaled PSF images were subtracted from the QSO to yield a
resulting
profile that increases monotonically to the nucleus. The value for
the profile that just turns over near the nucleus yields an upper
limit for the host galaxy flux. The best value was taken to be the
intermediate case where the difference flux follows a smooth profile for
its inner part. While this was subjective, and depends in principle
on whether the difference follows an exponential or de Vaucouleurs
profile, or combination, the QSO nuclear flux was generally small
enough that the spread in these difference fluxes was small. Error
bars were taken for each host galaxy flux as the upper and lower
values described above.

The same procedure was repeated for the co-added images for each QSO.
Table 1 shows the final adopted values for the ratio of nuclear to
resolved flux for each QSO. The agreement between the individual and
the combined images was in all cases within 20\%, and the ranges as
described above have similar values. Thus, in Table 1 we show the
mean values (average discarding outliers) for the nuclear to flux ratio, 
and we assign an
overall uncertainty of 30\% in those values. Note that the resolved flux
given does not include any separate close companions within the
radius of 10 arcsec: where these are present they were not included.
The resolved flux does include any features that appear to be part of
or connected to of the host galaxy, even though they introduce irregularities 
into the azimuthally averaged profile. Many of the profiles are
irregular, which means that fit to either standard model is
somewhat arbitrary. We have noted where the profiles do have some
region of good fit to one or other model. There are ten with some
spheroidal component and four with a good exponential. Four others
have so many companions that no fit is good, and another three are clean
but do not have clean regions of good fit. In several cases, the host
galaxy appears to have both bulge and disk components to the azimuthally
averaged profile. 

The morphology of the QSO images, both raw and PSF-subtracted, was
inspected to make an estimate of the degree of tidal disturbance.
As in paper 1, this value takes into account the presence of connecting
bridges to companions, single asymmetric arms or extensions, radial
jet-like features, warps, and other asymmetries. This is quantified as
an interaction strength index from 0 to 3, as in paper 1. Two objects have
archival HST images from Marble et al 2003 (the last and third last ones 
in Table 1). In the first of these, no interaction is seen in either
telescope image. In the second, our CFHT image has some asymmetry (interaction
index 1), but the HST image shows considerable structure that clearly
indicates a significant disturbance on small scales (see Figure 3). 
Thus, our interaction indices may well be conservative. Figure 3 shows 
examples of the three levels of interaction, plus an HST image for comparison.

\section{Results}

   All the QSOs in the sample are resolved, most of them easily, with 
ratio of nuclear to
resolved flux ranging from 0.17 to 25. We discuss the properties
of the resolved flux in the subsections below, but begin with a look
at possible biases and selection effects within the dataset.

\subsection{Biases and systematics}

   Since there is a range of image quality in the sample (from 0.6 to
1.3 arcsec), we looked for measured quantities that may depend on it.
The image quality shows no correlation with object redshift.
The interaction index shows no envelope or correlation. The ratio of
nuclear to resolved flux also does not show any trend, and certainly not
in the expected sense of lower ratios for better seeing. The dynamic range
of `useful' QSO image does show an upper envelope where the maximum range is
smaller for larger FWHM images. This is largely the effect of reducing the
peak central signal with poor image resolution, and the limiting signal is 
not affected. Overall, for the sample and measurements we discuss, the image
quality is not very important and is not introducing any significant bias
to the results.

   Another observational variable is the total exposure time. While most
objects have 680 secs integration, three have twice that, and two have
four times that. The interaction index, and useable dynamic range are
not correlated with the exposure, and the two best-resolved objects
have the minimum exposure. Thus, exposure time is not biassing the
overall results of the work.

  The mix of spectral types is not correlated with object redshift. The
magnitude of the QSO is not correlated with the dynamic range of the
images. The least resolved objects are not the highest redshift ones, so
there is no observational bias obvious. The median ratio of nuclear to
resolved flux is 4, and values near this are spread evenly across the
redshift range. There is no systematic colour change with redshift, or
change of scale length with redshift, so the mix of objects appears to
be unrelated to redshift.

\subsection{Nucleus to host ratio}

   Figure 4 shows the ratio of nuclear to resolved flux with spectral
class. The ratio increases with the dominance of the broad emission
line components, which is as expected for the general model of
central obscuration by a torus for type 2 objects.

   The lowest dynamic range observations have the lowest nuclear light
domination. This means that the resolved flux is in the inner parts
of the host galaxies, and that low nuclear domination is caused by
obscuration within
the central host galaxy. However, overall QSO colour is loosely
correlated with nuclear domination, so the obscuration is connected with
reddening of the nuclear light - a separate phenomenon from the
obscuration of line emission by the torus.

The result that obscuration is connected with the diminishing
of the nuclear light in the optical is supported by the findings
of Francis, Nelson and Cutri (2004) who found that a number of
near-ir flux-selected AGN were missed in SDSS because their optical colors
were indistinguishable from normal galaxies.  They speculated this
was because the nuclei were preferentially reddened and thus better
visible in the near-ir.

   Scale length of the host galaxy is not correlated with nuclear domination,
so the nuclear region is obscured without affecting the outer parts of the
host galaxy. Scale length is also not related to the spectral class or the
colour of the QSO. Further mention is made of the scale length below.

\subsection{Interaction status}

  The level of interaction seen is not strongly related with the QSO colour,
but there is a trend towards redder colour for more strongly interacting
systems. There is a clear dependence and upper envelope with the nuclear
fraction, in the sense that the more interacting objects have more 
obscured nuclei (see Figure 5).

   It is also clear that interactions are less obvious in higher redshift
objects, as expected as the signal to noise and the angular scales decrease.
We note that the object with HST imaging, given interaction class 1 from
the CFHT data, is clearly interacting in the HST images, and would have
index 2. Thus, the CFHT interaction indices should be regarded as lower limits
overall, and particularly for the higher redshift objects.

   The host galaxy scale length shows a trend where they are larger
for weaker interacting systems, although there is considerable spread.
This will reflect the presence and decay of disks, tidal arms, and
eventual increase in spheroid structure during a major interaction.

   Comparison with the z$<$0.3 sample in paper 1 is of interest. Figure 6
shows the fractions with interaction index with redshift in different
QSO samples,
where the 2MASS lower redshift sample is large enough to split at
z=0.2. There is a
systematic shift towards lower interaction index with increasing redshift.
Even allowing for lowering the index by one grade point for 1/2 of them
(which would allow for missed signatures seen only with higher resolution,
as is Figure 3),
we find the interactions are stronger at lower redshifts. Since the
higher redshift objects are more luminous, it may be that the more interacting
and hence more obscured objects are not detected in the higher redshift
bins because of flux limits. The ratio of counts of objects above and below
z=0.3 (64 to 243, plus lack of LINERS at z$>$0.3), shows the 2MASS flux
limitation clearly.

\subsection{Radio flux}

  Six of the sample of 21 appear in the FIRST radio catalogue (although
2 of them
are too far south to appear in the FIRST catalogue: see Becker et al 1995).
They are
all unresolved, and faint - see Table 1 for the fluxes. The mean redshift
of the radio sources is the same as that for the others ($\sim$0.40).
There is no correlation between radio flux and colour. The mean ratio
of nucleus to host for the radio sources is 2.5, compared with 7.2 for
the others (medians 1.3 and 3.2). The spectroscopic types have the same
(full) spread for radio and radio-quiet objects. The radio sources
have an average interaction index of 2 while the others have value 1,
and this difference is significant at the 95\% level, if the
distributions
are gaussian. This suggests that
nuclear radio sources occur in recently activated nuclei where the
signs of interaction, and dust obscuration are higher. This is similar
to the result on IRAS galaxies investigated by Neff and Hutchings
(1992).

\subsection{Asymmetries and and profiles}

   The ellipticities of the contours in this sample are lower than
those in paper 1, and follow the same upper limit which decreases with
increasing redshift. This is largely the effect of the diminishing scale and
surface brightness with redshift, but the values are less determinate as the 
presence of line of sight companions increases too.

   Profiles were classified as good fits to R$^{1/4}$ law or
exponentials
if they had significant linear sections in the relevant plots. Many
objects have nearby companions or asymmetries which made these simple
classifications impossible. However, there were 3 objects with good
exponential
profile sections and 9 with good bulge components. Their mean redshifts
are 0.38 and 0.43, respectively, so that the outer exponential tails may
simply be less detectable in the fainter higher redshift objects.

   It is of interest to compare the profiles of an object with HST imaging
(1715+281).
The HST image shows strong asymmetry of the resolved flux and is traceable to
a radius of 2.4", while the CFHT image shows less of the structure but
the same asymmetry out to 7" (see Figure 2). 
The profiles agree and both indicate
that r$^{1/4}$ fits quite well, while an exponential does not. Thus, there
is good agreement in this one case of overlap with HST data.

\subsection{Scale lengths}

   The scale lengths given in Table 1 are derived from the slopes of the
radius-magnitude plots for the images, outside the unresolved cores. They
do not include any signficant companions, and are converted to Kpc for
a Hubble constant of 75. Cases where the slope cannot be measured
with any reliability are left blank. Scale length is not correlated
with the dynamic range of the images, although the latter range only
by about 25\%. It is also not correlated with the image quality - as
expected
since the profiles are azimuthally averaged over values far greater than
the image quality.

  The scale length is larger for the more interacting systems. This
is a quantitative measure of the extended arms and asymmetries that
lead to the higher interaction indices. In general the scale length values
are comparable with large nearby galaxies.

\section{Discussion}

    In paper 1 we noted a number of differences between the 2MASS low
redshift QSOs and standard blue QSOs. Generally speaking, the 2MASS obects
are redder and have more obscured nuclei, and the host galaxies show a far
greater proportion of tidal interactions.

   We note that the 2MASS sample requires detection in all three of the
JHK passbands, so that there is a bias against highly reddened objects.
The J-K colours average at 2.2 mag in both the lower and higher
redshift subsamples, but the B-R colours are different, as seen in Table 2.
The selection will also lose the less luminous objects as the redshift
increases. While this is clearly true from Figure 1, it is interesting to 
note the large fraction of more luminous objects that appear at higher
redshifts. This increase is just what is expected from the 6-fold increase 
in volume of space sampled, so there may not be significant selection effects
between the present sample and the lower redshift objects in paper 1.

    In most of the aspects considered, the higher redshift 2MASS objects are
more similar to `normal' blue-selected QSOs, but they are intermediate between
them and the low redshift 2MASS QSOs. Table 2 shows some key comparison
numbers, based on paper 1 and the optically selected sample of Hutchings
and Neff (1991).

    The higher redshift sample of this paper correspond to the highest
luminosity objects in paper 1. The subset of the paper 1 objects that
match this K-band luminosity are also shown in Table 2, and they have higher
nucleus to host ratio and bluer colour than the full paper 1 sample,
but not as high or blue as the sample in this paper. Thus, luminosity
is a relevant parameter as well as redshift. The single high redshift object
(at z=2.37) has very high luminosity (although optical emisssion lines 
shifted into the 2MASS bandpasses probably contribute), and is hardly
resolved in our data. Thus, there is probably nothing very remarkable
in this object from the data presented here.

   The higher redshift (and luminosity) sample of this paper is
generally similar to normal blue QSOs, but still have a higher fraction of
interaction evidence, although the fraction of highly interacting objects
is similar to that for blue QSOs. This is seen in both morphology and
luminosity profiles. It seems
likely that the higher luminosity QSOs blow away the circumnuclear dust
faster, but are still relatively young AGN. The unresolved nature of the 
radio sources is consistent with this idea.

   The sample does not reach the faint luminosities of the low redshift
objects in paper 1, so the evolution of these sources cannot be
traced until we have a deeper NIR QSO survey. We note that the Spitzer
telescope has reported finding a large population of lower luminosity
AGN in their deeper survey. It will be important to follow those up
with high resolution imaging. 

This publication makes use of data products from the Two Micron All Sky
Survey, which is a joint project of the University of Massachusetts and the
Infrared Processing and Analysis Center/California Institute of Technology,
funded by the National Aeronautics and Space Administration and the
National Science Foundation.

\newpage
\scriptsize
\begin{deluxetable}{lrllrrccrlrc}
\tablenum{1}
\tablecaption{2MASS QSO sample}
\tablehead{\colhead{RA} &\colhead{Dec}
&\colhead{z} &\colhead{Typ} &\colhead{B, R, J, H, K} &\colhead{N/H}
&\colhead{Int} &\colhead{Sc L} &\colhead{Range} &\colhead{20cm}
&\colhead{IQ} &\colhead{M$_K$}\\
&&&\colhead{\tablenotemark{a}} &&\colhead{\tablenotemark{b}}
&\colhead{\tablenotemark{c}} &\colhead{Kpc\tablenotemark{d}}
&\colhead{mag\tablenotemark{e}} &\colhead{mJy} &\colhead{"\tablenotemark{f}}}
\startdata
11 03 12.93  & 41 41 54.9  &0.403 &1.2
&16.6,16.1,15.2,14.2,13.0 &16.6&1. &-- &9.5&-- &1.1 &-29.1\\
13 32 31.17  & 03 59 28.0  &0.346& 1.
&17.4,17.3,16.0,14.8,13.4 & 4.3& 1.&--& 11.3&1.5
& 0.9 &-28.3\\
13 45 17.89  & -08 29 57.3 &0.473& 1.
& --,  ~~~~--, 15.6,14.2,12.8
& 3.9& 2.&7.7 &9.6&-- & 0.8 &-29.6\\
14 32 04.62  & 39 44 38.9 &0.349&1.5
&16.4,16.3,16.0,15.4,14.4
& 5.4& 1. &4.4 &8.5&-- & 0.8 &-27.3\\
14 35 15.66  & 02 32 21.7 &0.305&1.2
&16.6,16.7,15.6,14.4,12.9
& 12.& 1.&--& 11.5&-- & 0.9 &-28.5\\
14 38 27.94  & -11 22 49.5 &0.401&1.5
&19.6,18.7,16.3,15.1,13.6
&1.2& 1.&8.6 & 9.0 &-- & 1.1 &-28.4\\
14 41 18.87  & -11 31 47.5 & 0.330 &1.2
&16.6,16.8,16.2,15.2,14.0
& 2.5& 1.&4.8 & 10.5&-- &1.0 &-27.6\\
14 42 02.95  & 14 55 39.3 &0.307&1.2
&17.6,17.0,16.2,15.2,14.2
& 1.8& 3. &5.9 & 10.4& 104. & 0.9 &-27.2\\
14 50 00.90  & 14 29 48.7 &0.358&1.2
&17.7,17.1,16.4,15.6,14.4
& 12.& 0.&--&12.0 &-- & 0.7 &-27.4\\
15 00 13.40  & 12 36 45.2 &0.407& 1.9
&19.5,18.9,16.8,15.9,14.6
& 0.3& 3. &4.7 &8.1&-- & 0.8 &-27.5\\
15 01 50.51  & 49 33 38.2 &0.337&1.9
&18.1,17.1,16.4,14.9,13.5
&0.26& 3.&4.8 & 8.0 &3.1 & 0.8 &-28.2\\
15 31 07.19  & 12 08 14.4 &0.542&1.5
&19.8,18.7,17.1,15.7,14.6
& 3.2& 2.&6.7& 10.2&-- &1.0 &-28.1\\
15 36 44.92  & 14 12 29.4 &0.399&1.2
&16.7,17.3,16.1,15.4,14.0
& 4.8& 2. &6.6 & 10.5&-- & 0.9 &-28.0\\
15 40 19.57  & -02 05 05.3 &0.319& 1.
& 16.0,15.8,15.3,14.3,13.2
&6.& 2.&4.7 &9.3&4.7 &1.0 &-28.3\\
15 49 38.73  & 12 45 09.2 & 2.37& 1.9
&18.7,17.3,15.8,14.5,13.5
& 25.& 0. &-- &12.5 &-- & 0.7 &-33.0\\
15 50 59.30  & 21 28 08.8 &0.373& 1.
&17.3,16.8,16.3,15.7,14.3
& 8.8& 0.&--& 12.8&-- & 0.6 &-27.6\\
16 18 09.74  & 35 02 08.9 &0.446& 1.9
&18.8,18.2,16.8,15.4,14.1
& 2.6& 2.&6.9&10.0 &14. & 0.9 &-28.2\\
16 44 20.14  & 56 36 44.6 &0.329&1.5
&19.8,18.2,17.2,15.9,14.6
&0.17& 1.&4.7 &9.5&-- & 1.3 &-27.0\\
17 00 02.99  & 21 18 23.3  &0.596 &1.5
& --,  ~~~~--, 17.4,15.9,14.9
&1.4 & 0.&3.8 &8.3&-- & 1.1 &-28.1\\
17 00 56.01  & 24 39 28.2 &0.509&1.5
&16.8,16.8,16.0,15.3,14.3
& 10.& 1. &8.7 & 10.3&-- & 0.7 &-28.3\\
17 15 59.77  & 28 07 16.9  &0.524 &1.8
& --,  ~~~~--, 17.2,15.8,14.6
&0.32& 1. &4.4 &7.8&1.6 & 0.8 &-28.0\\
\enddata
\tablenotetext{a}{ The range from 1 to 2 reflects the ratio of narrow to
broad emission lines.}
\tablenotetext{b}{ The ratio of nuclear to host flux in r-band, from
this work. Uncertainties are estimated to be 30\%.}
\tablenotetext{c}{ Strength of host galaxy interaction based on observed
morphological features, as in paper 1.}
\tablenotetext{d}{ 1/e profile flux drop}
\tablenotetext{e}{ Total dynamic range of profile in magnitudes}
\tablenotetext{f}{FWHM of stellar images in arcsec}

\end{deluxetable}

\newpage
\normalsize
\begin{deluxetable}{lcccc}
\tablenum{2}
\tablecaption{2MASS QSO sample comparisons}
\tablehead{\colhead{Property} &\multicolumn{3}{c}{2MASS QSOs}
&\colhead{Blue QSOs}\\
&\colhead{z$<$0.3} &\colhead{z$<$0.3,luminous} &\colhead{z$>$0.3} }
\startdata
Average z &0.2 &0.2 &0.4 &0.2\\
Sample size &76 &12 &21 &28\\
\\
B-R &1.1 &0.8 &0.5 &0.3\\
Nuc/Host &0.5 &1.2 &6.0 &5.0\\
Int strong &33\% &42\%&14\% &11\%\\
Int (any) &75\% &58\% &81\% &39\%\\
Exp profile &16\% &-- &14\% &18\%\\
Bulge profile &20\% &-- &48\% &54\%\\
Messy profile &64\% &-- &38\% &28\%\\
Radio detected &40\% &50\% &29\% &43\%\\
\enddata
\end{deluxetable}

\newpage
\centerline{References}

Becker, R., White R.L., and Helfand D.J., 1995, ApJ, 450, 559

Cutri,R.M. et al, 2002, ASP Conf Ser vol 284, 127

Francis P.J., Nelson B.O., and Cutri, R.M., 2004, AJ, 127, 646

Hutchings J.B., Maddox N., Cutri R.M., Nelson B.O., 2003, AJ, 126, 63
(paper 1)

Hutchings J.B. and Neff S.G., 1992, AJ, 101, 434

Marble A.R. et al., 2003, ApJ, 590, 707

Neff S.G. and Hutchings J.B., 1992, AJ, 103, 1992

\newpage
\normalsize
\centerline{Captions to figures}

1. Sample selection in this paper (z$>$0.3) and paper 1, from the full
2MASS sample observable from CFHT. Filled circles are the objects observed. 
Top: dashed lines are contours of constant luminosity. Bottom: spectral
types range from broad to narrow emission lines, with intermediate
values. Dashes are the full sample and dots are those observed.
The paper 1 observed sample is grouped as types 1 and 2 only, as given
in that paper.

2. Azumuthally averaged profiles of QSO 1715+281, with the CFHT PSF 
and also an HST image (with slightly different filter).

3. Images of representative objects from the sample, in decreasing levels
of interaction - top left (1500+12) level 3, top right (1442+14) level 2, 
bottom (1715+28) level 1. The lower right is the HST image of the object 
at lower left. The images are 11 arcsec on a side except for the HST which
is 6 arcsec.

4. Correlations with nuclear light fraction. Top: the type 1 objects
have higher nuclear flux, consistent with the orientation expectation
for broad-line objects. Bottom: With the high value exception, there
is correlation with overall QSO colour, consistent with the nuclear
fraction being reduced by dust reddening. In the lower panel, open dots 
are B-K and filled dots are 3(B-R). The line is the linear fit 
through all except the top right object.

5. Correlations with interaction level of host galaxies. The interaction
scale was estmated on a 5-point scale but is reduced to 3 here to match
the values from paper 1. Top: high
levels of interaction are not seen in the highest redshift objects,
presumably because of signal level and angular scale. Bottom: Highly
interacting hosts have low flux nuclei, presumably because they are
obscured by dust.

6. Fraction of hosts in different interaction levels in three redshift
bins with about equal sample numbers. There is a systematic change with
redshift whereby lower redshift objects are more interacting. This may
not all be due to detectability changes with redshift.

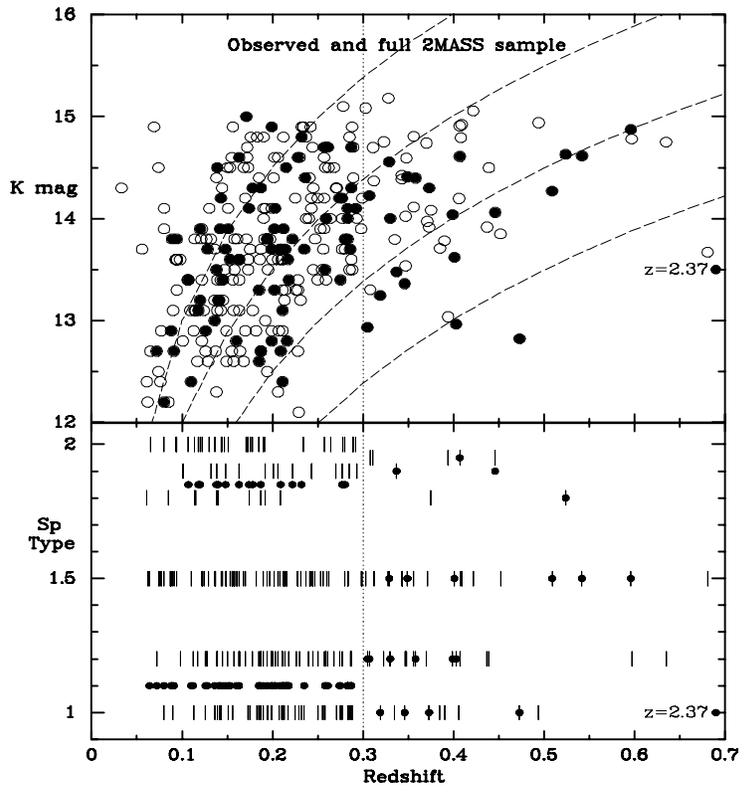
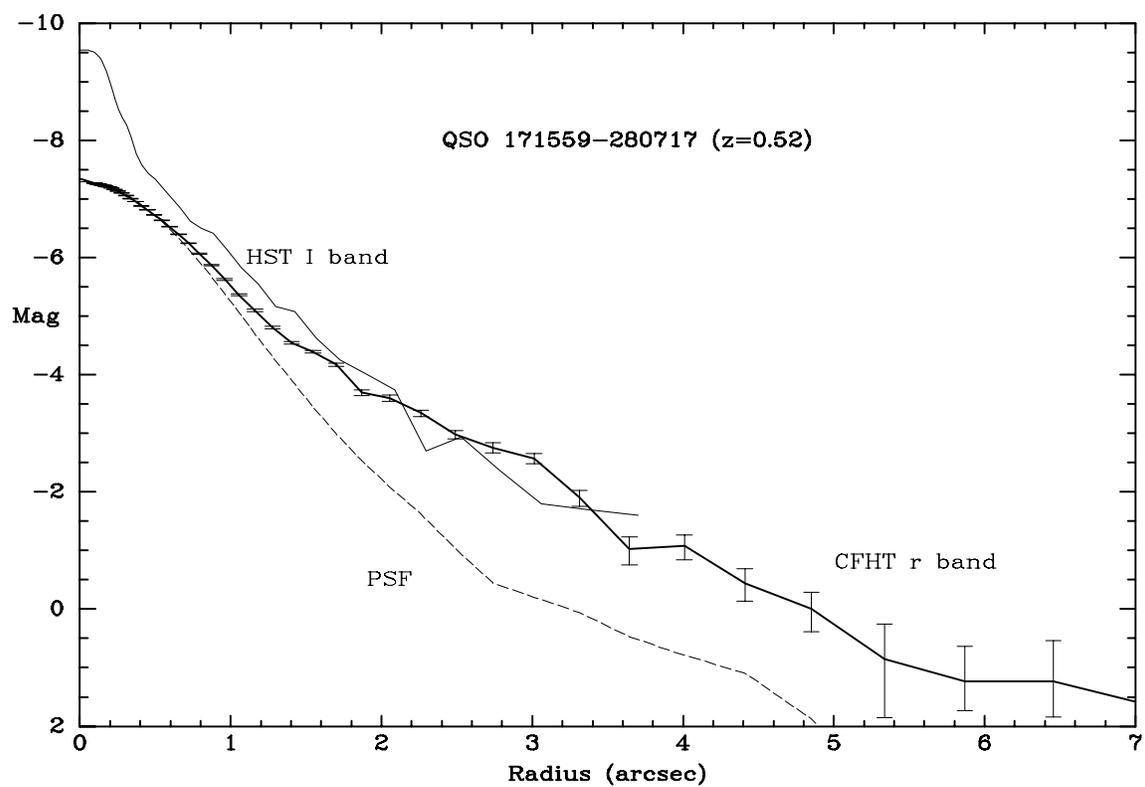
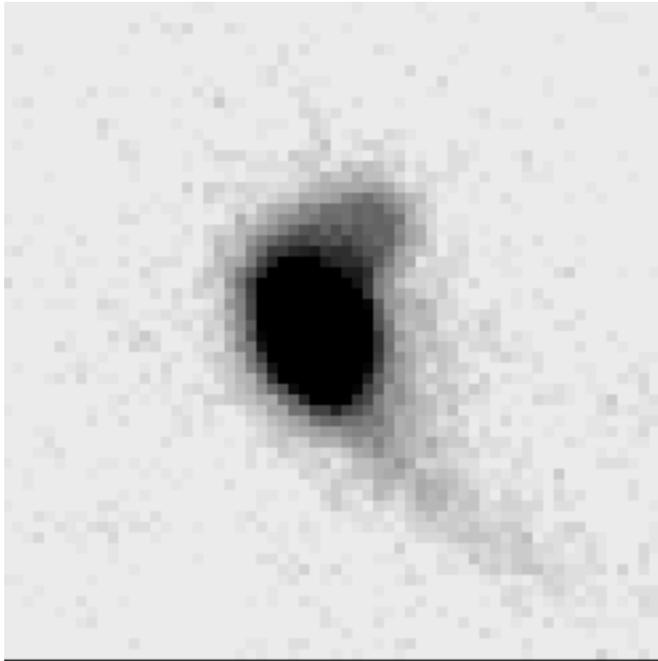
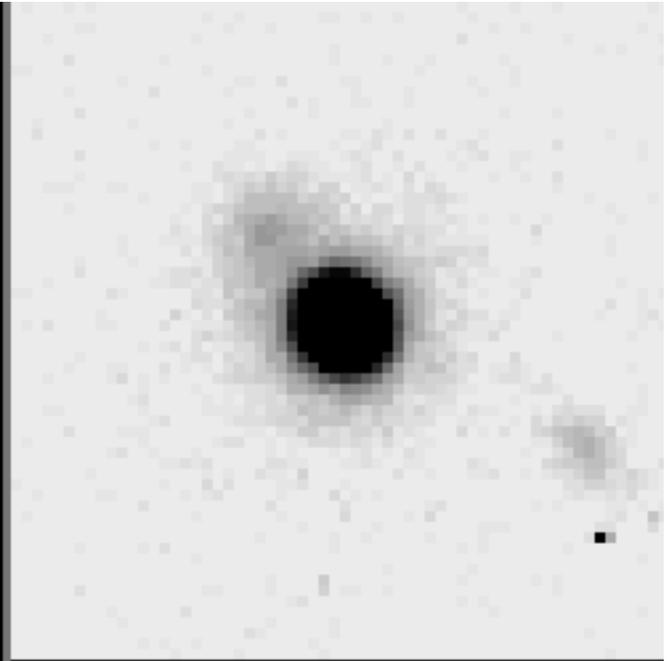
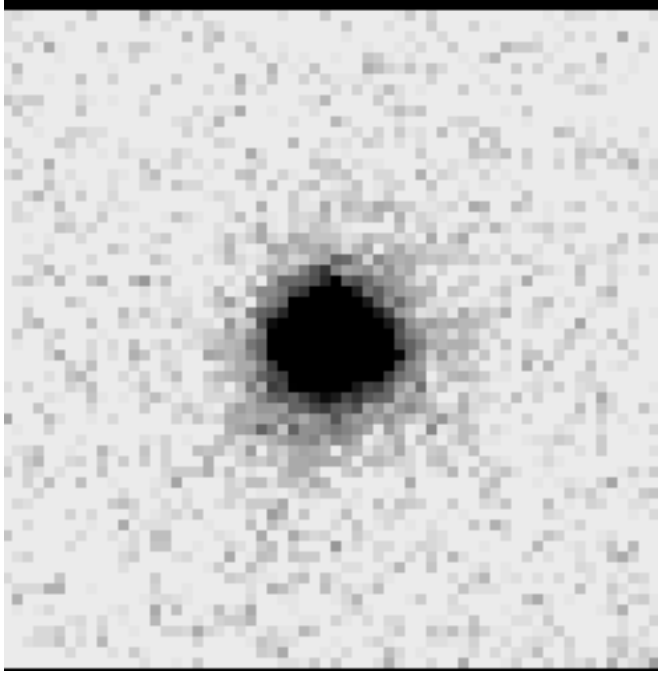
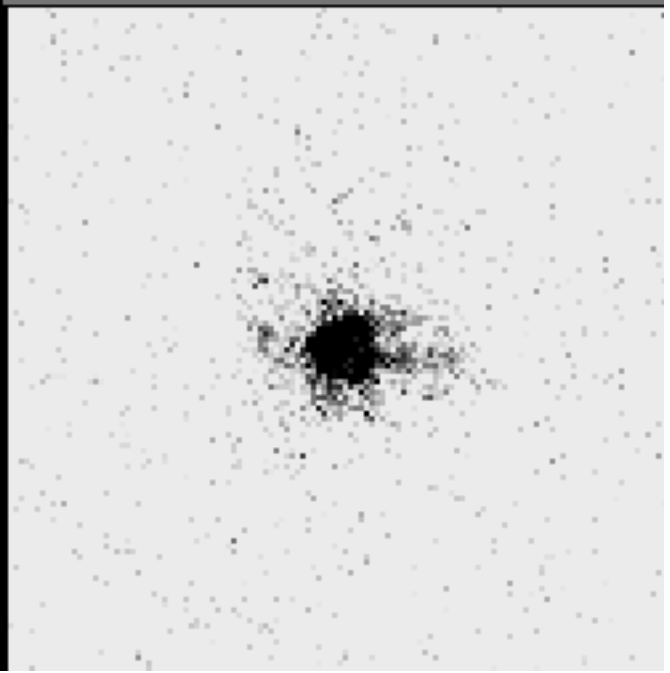
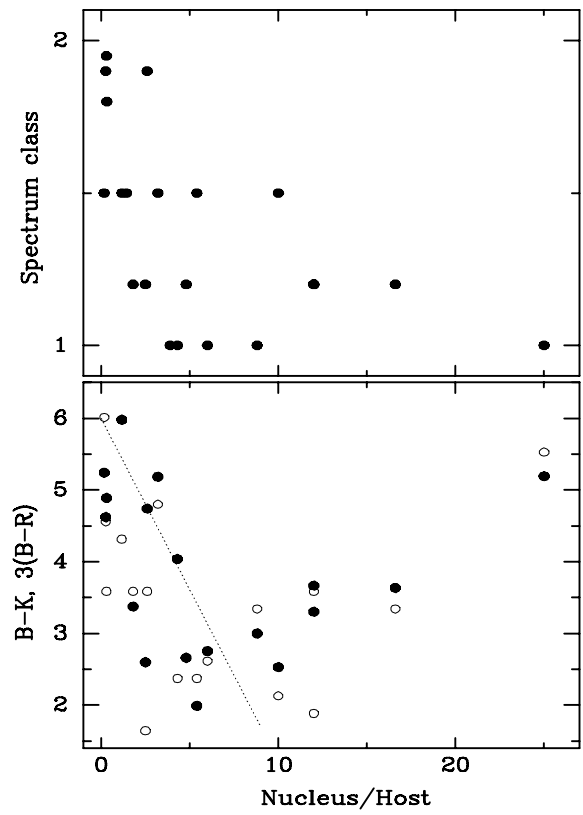
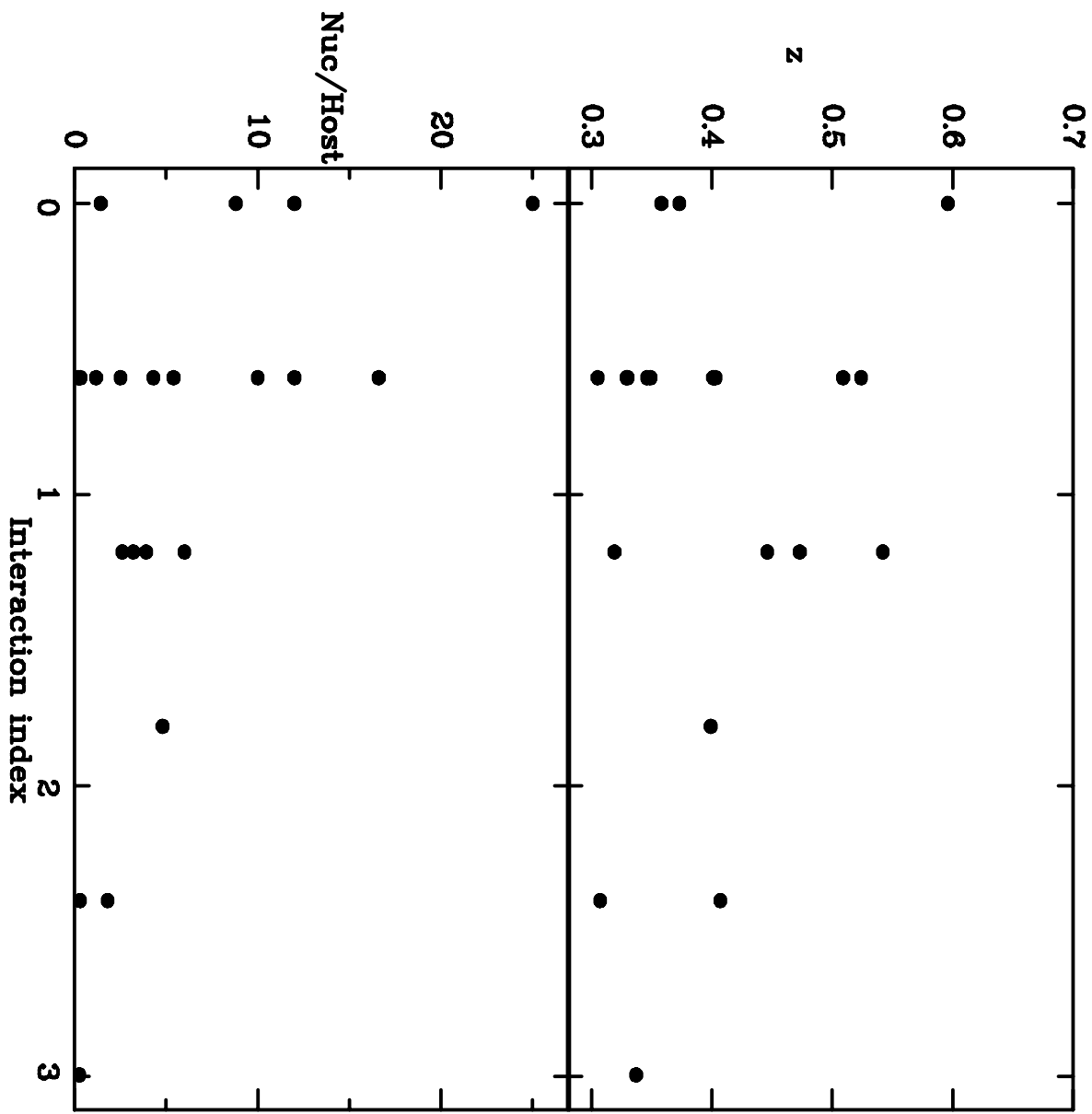
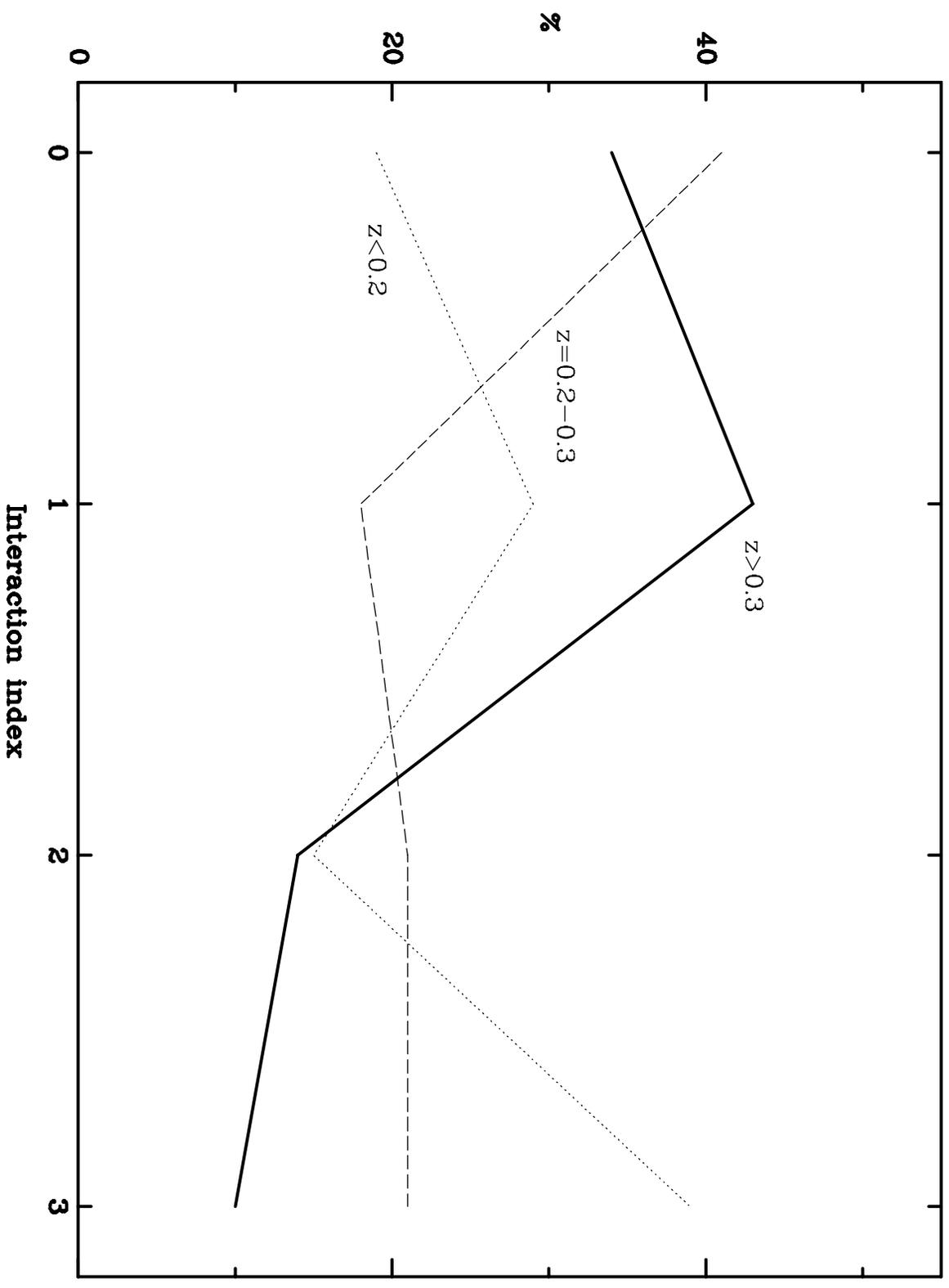
\end{document}